# Detecting Single Infrared Photons with 93% System Efficiency

F. Marsili[1*], V. B. Verma[1], J. A. Stern[2], S. Harrington[1], A. E. Lita[1], T. Gerrits[1], I. Vayshenker[1], B. Baek[1], M. D. Shaw[2], R. P. Mirin[1], and S. W. Nam[1]

[1]*National Institute of Standards and Technology, 325 Broadway, MC 815.04, Boulder, CO 80305, USA*

[2]*Jet Propulsion Laboratory, 4800 Oak Grove Dr., Pasadena, California 91109, USA*

[*]*corresponding author: francesco.marsili@nist.gov*

**Introductory paragraph**

Single-photon detectors (SPDs)[1] at near-infrared wavelengths with high system detection efficiency ($> 90\%$), low dark count rate ($< 1$ counts per second, cps), low timing jitter ($< 100$ ps), and short reset time ($< 100$ ns) would enable landmark experiments in a variety of fields[2-6]. Although some of the existing approaches to single-photon detection fulfill one or two of the above specifications[1], to date no detector has met all of the specifications simultaneously. Here we report on a fiber-coupled single-photon-detection system employing superconducting nanowire single-photon detectors (SNSPDs)[7] that closely approaches the ideal performance of SPDs. Our detector system has a system detection efficiency (*SDE*), including optical-coupling losses, greater than 90% in the wavelength range $\lambda = 1520 - 1610$ nm; device dark count rate (measured with the device shielded from room-temperature blackbody radiation) of $\approx 0.01$ cps; timing jitter of $\approx 150$ ps FWHM; and reset time of 40 ns.

**Letter**

Superconducting nanowire single photon detectors (SNSPDs)[7] are based on thin (thickness below 10 nm) and narrow (width below 200 nm) superconducting nanowires current-biased near the critical current. The absorption of one or more photons in a superconducting nanowire drives part of the nanowire to the normal state with a resistance of the order of several kilohms[8-10], which can be detected by using an appropriate readout circuit[7]. SNSPDs have outperformed other near-infrared single photon detector technologies in terms of dark count rate, timing resolution and reset time [1]. However, after over ten years of research, to date the system detection efficiency (*SDE*, which includes the efficiency of the optical coupling to the detector) of SNSPDs has been limited to ~ 20%[11,12] at 1550 nm wavelength ($\lambda$), due to the limited compatibility between the superconducting material used (typically, NbN) and the structures to enhance the *SDE*. Because the superconducting properties of NbN films depend on the crystal phase of the films[13] and are affected by crystal defects[14,15], using NbN nanowires makes it challenging to meet the two requirements to achieve high-*SDE* SNSPDs: (1) the superconducting properties of nanowires and of bulk samples are similar, which ensures the correct operation of the detector; (2) the light is coupled to the detector and absorbed in the nanowires with high efficiency. The fragility of superconducting NbN with respect to structural disorder limits: (1) the fabrication yield of large-area devices[16], (2) the choice of substrates for fabrication, and (3) the design parameters of optical structures that would enhance the absorption in the nanowires. We recently reported on the fabrication of SNSPDs based on a different superconducting material, amorphous tungsten silicide ($W_{0.75}Si_{0.25}$, or WSi)[17]. Because the crystal structure of WSi is homogeneously disordered, WSi superconducting nanowires are more robust with respect to structural defects than NbN nanowires (which allows the fabrication of larger area devices), can be deposited on a variety of substrates (we have already successfully fabricated WSi SNSPDs on thermally grown $SiO_2$[17] and on a spin-on glass[18]), and allow more degrees of freedom in optimizing the optical coupling to the detectors. Therefore, WSi appears as a strong candidate to achieve high-*SDE* SNSPDs. Here we report on WSi SNSPDs embedded in an optical stack designed to enhance absorption (see Supplementary Information) at $\lambda = 1550$ nm and coupled to single-mode optical fibers at $\lambda = 1550$ nm with a self-aligned mounting scheme based on Si micromachining[19]. Our WSi SNSPDs achieved a system detection efficiency (*SDE*) as high as ~ 93% around $\lambda = 1550$ nm; system dark count rate of ~ 1 kcps (due to the room-temperature blackbody radiation); timing jitter of ~ 150 ps full width at half maximum (FWHM); and reset time of 40 ns.

We characterized our single photon detection system (see Methods and Supplementary Information) by employing 16 different SNSPDs from four fabrication runs. The devices were mounted in a cryogen-free adiabatic demagnetization refrigerator and operated in the temperature range $T = 120$ mK - 2 K. The response pulse of the detector was read-out with room-temperature electronics.

Figure 1a shows the system detection efficiency (*SDE*, see Methods and Supplementary Information) as a function of the bias current ($I_\text{B}$) for our best device measured with 1550 nm light. As the detection efficiency of SNSPDs varies with the polarization of the incident light[20], the polarization state of the light was varied on the Poincaré sphere to maximize or minimize the counts from the detector. We therefore obtained a maximum ($SDE_\text{max}$, red curve) and minimum ($SDE_\text{min}$, blue curve) *SDE* vs $I_\text{B}$ curve. Both the $SDE_\text{max}$ and $SDE_\text{min}$ curves showed a sigmoidal shape and saturated at $SDE_\text{max} \approx 93\%$ and $SDE_\text{min} \approx 80\%$ for $I_\text{B}$ larger than a cut-off current $I_\text{co} = 1.5$ μA (the bias current at the inflection point of the *SDE* vs $I_\text{B}$ curve[21], see Figure 1a) and lower than the switching current of the device $I_\text{SW} = 4$ μA (the maximum current the device could biased at without switching to the normal, non-superconducting, state, see Figure 1a). At $I_\text{B} = 3$ μA, the average and (one-sigma) uncertainty of the maximum and minimum *SDE* were (see Supplementary Information) $SDE_\text{max} = 93.2 \pm 0.4\%$ (red circle in Figure 1a) and $SDE_\text{min} = 80.5 \pm 0.4\%$ (blue circle in Figure 1a). The experimental value of the *SDE* was lower than the design value of the absorption of the SNSPDs (> 99%), which we attribute to several possible causes: (1) our imperfect knowledge of the refractive index of the materials used in the optical stack; (2) fabrication imperfections; (3) coupling losses; and (4) the non-unity internal detection efficiency of the SNSPDs (the probability that the absorption of one photon in a nanowire results in a response pulse).

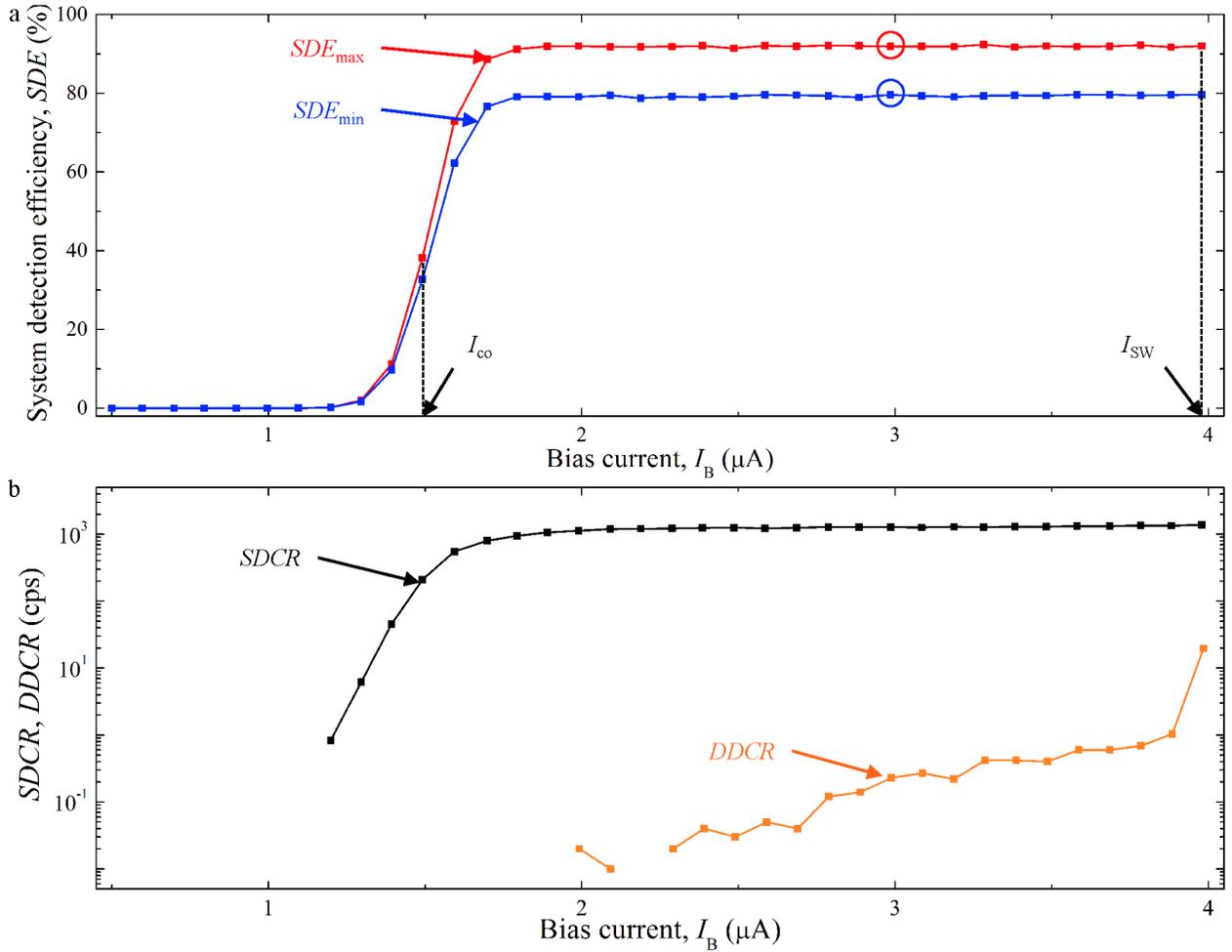

**Figure 1. Bias-current dependence of the system detection efficiency ($SDE$), the system dark count rate ($SDCR$), and the device dark count rate ($DDCR$). a.** $SDE$ vs bias current ($I_B$) for two different polarizations of the light at $\lambda = 1550$ nm and $T = 120$ mK. The SNSPD employed was based on 4.5-nm-thick, 120-nm-wide nanowires with 200-nm pitch. The SNSPD covered a square area of 15 μm × 15 μm. The dashed lines indicate the $I_{co}$ and the $I_{SW}$ of the device. **b.** $SDCR$ and $DDCR$ vs $I_B$ for the same device of Figure a. The $SDE_{max}$, $SDE_{min}$ and $SDCR$ curves were obtained by averaging 6 subsequent

acquisitions of the curves. The error bars for each point are not plotted for clarity, but the uncertainty is described in Supplementary Information.

Figure 1b shows the bias dependence of the system dark count rate (*SDCR*, the response pulse count rate measured when the input fiber to the system was blocked by a shutter) and of the intrinsic dark count rate of the SNSPD, or device dark count rate (*DDCR*, the response pulse count rate measured when the fiber was disconnected from the device inside the refrigerator). The *SDCR* vs $I_B$ curve had a sigmoidal shape similar to the *SDE* vs $I_B$ curves shown in Figure 1a and saturated at $SDCR \approx 1$ kcps for $I_B > I_{co}$. The *DDCR* was $< 1$ cps for most of the bias range, which was approximately two-orders-of-magnitude lower than the *DDCR* of NbN SNSPDs of similar active area and fill factor[22]. Therefore, we concluded that the *SDCR* was dominated by blackbody photons[23].

Usually, the detection efficiency of SNSPDs varies significantly with the polarization of the incident light (typically by a factor of $\approx 2$[12,20]). However, a detector with polarization-insensitive *SDE* would be desirable for many applications[1]. Therefore, we characterized the polarization and wavelength dependence of the *SDE* by mapping the *SDE* onto the Poincaré sphere in the wavelength range $\lambda = 1510 - 1630$ nm. Figure 2a and 2b show the *SDE* (in color scale) as a function of the inclination ($2\theta_P$) and azimuth ($2\varepsilon_B$) angles of the polarization vector on the Poincaré sphere (we call this plot a *Poincaré map* of the *SDE*) at $\lambda = 1510$ nm (Figure 2a) and $\lambda = 1625$ nm (Figure 2b). Although the positions of the maxima and minima of the Poincaré maps were approximately the same at the two wavelengths, the ratio ($R = SDE_{max} / SDE_{min}$) between maximum and minimum values of the *SDE* changed from $R = 1.23$ at $\lambda = 1510$ nm to $R = 1.08$ at $\lambda = 1625$ nm. We obtained the wavelength dependence of $SDE_{max}$ and $SDE_{min}$ by extracting the maximum and minimum of the Poincaré maps at each wavelength. Figure 2c shows the wavelength dependence of $SDE_{max}$ (red squares), of $SDE_{min}$ (blue squares) and of $R$ (black triangles). The values of $R$ measured for $\lambda = 1510 - 1630$ nm were significantly lower than the value ($R \approx 2$) reported in Ref. [12,20], due to the relatively thin ($\approx 670$ nm total thickness) optical stack of materials used to increase the optical absorption in the WSi.

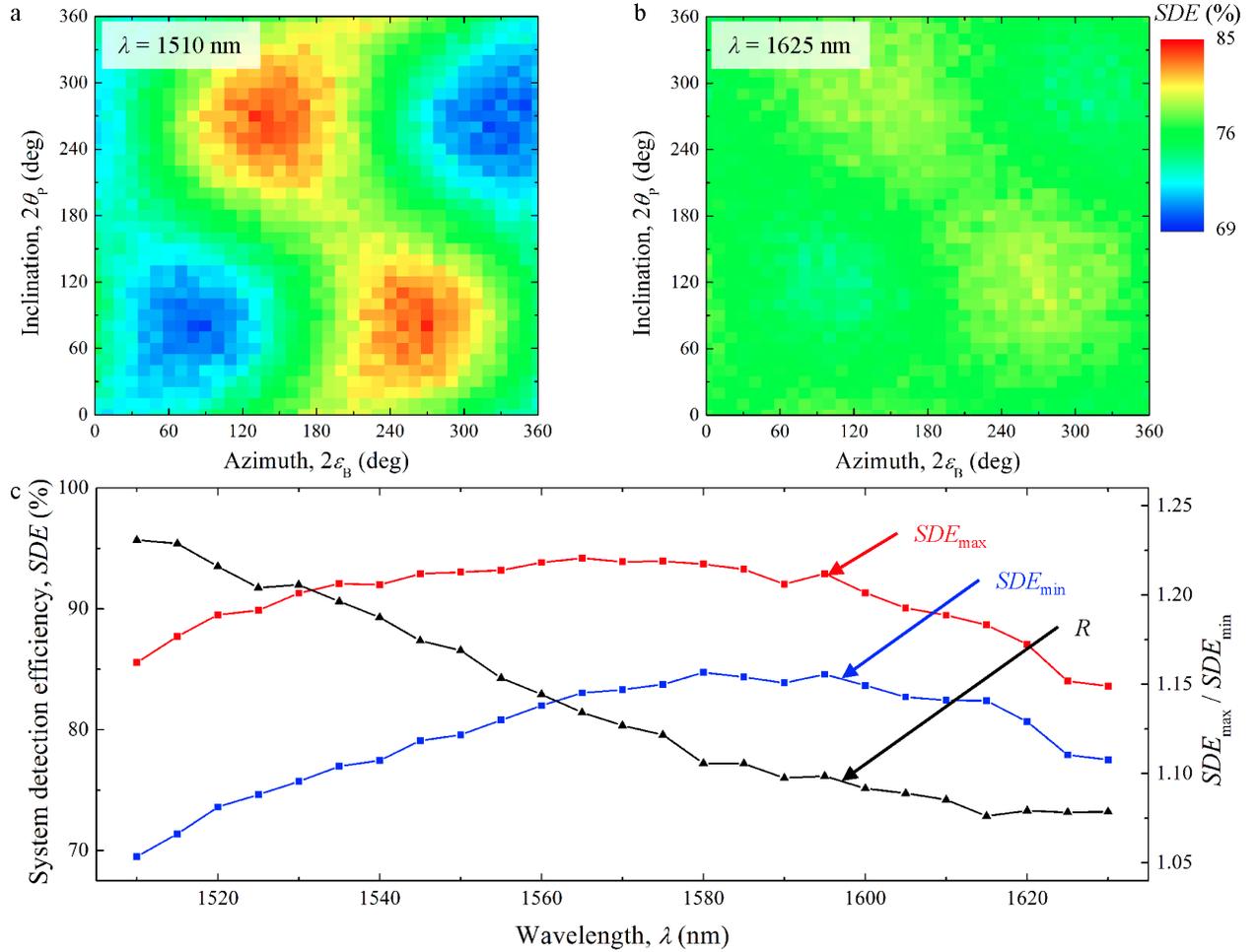

**Figure 2. Polarization and wavelength dependence of the *SDE*. a, b.** *SDE* (in color scale) vs the inclination ($2\theta_P$) and azimuth ($2\varepsilon_B$) angles of the polarization vector on the Poincaré sphere at (a) $\lambda = 1510$ nm and (b) $\lambda = 1625$ nm. The bias current was $I_B = 3.8$ μA. **c.** Wavelength dependence of the maximum *SDE* (*SDE*$_{max}$), of the minimum *SDE* (*SDE*$_{min}$) and of the ratio $R = SDE_{max} / SDE_{min}$. The measured parameters of the optical stack were, from top (illumination side) to bottom: 213 nm-thick TiO$_2$; 231 nm-thick SiO$_2$; 4.5 nm-thick, 120 nm-wide WSi nanowires with 200 nm pitch; 230 nm-thick SiO$_2$; 80 nm-thick Au. The width and the pitch of the nanowires were measured by scanning electron microscope. The experimental *SDE*$_{max}$ vs $\lambda$ and *SDE*$_{min}$ vs $\lambda$ curves were obtained by averaging 3 subsequent acquisitions.

Most of the readily accessible closed-cycle refrigeration technologies[24] do not reach a base temperature below 1 K. Therefore, it would be desirable to operate our detector above 1 K without degrading its performance. As the critical temperature of our SNSPD was $T_C = 3.7$ K, we characterized the performance of the system as a function of temperature by measuring the *SDE*, the *SDCR* and the *DDCR* in the temperature range $T = 120$ mK – 2 K. As shown in Figure 3a, the available bias range of the detector decreased with increasing $T$ because $I_{SW}$ approached $I_{co}$. However, the *SDE* at $\lambda = 1550$ nm remained saturated at $\approx 93\%$ for $T \leq 2$ K. The *DDCR* at the switching current ($I_B \approx I_{SW}$) increased with temperature from $\approx 20$ cps at $T = 120$ mK to $\approx 10$ kcps at $T = 2$ K. As shown in Figure 3b, for $T > 0.8$ K and $I_B > 0.95 I_{SW}$ the *DDCR* dominated the *SDCR*. Although the bias range for efficient, low-dark-count-rate single-photon detection decreased with increasing the temperature, the detector showed *SDE* $\approx 90\%$ and *DDCR* $< 10$ cps up to $T = 2$ K, confirming that we could operate the detector system at relatively high cryogenic temperature without significantly degrading its performance.

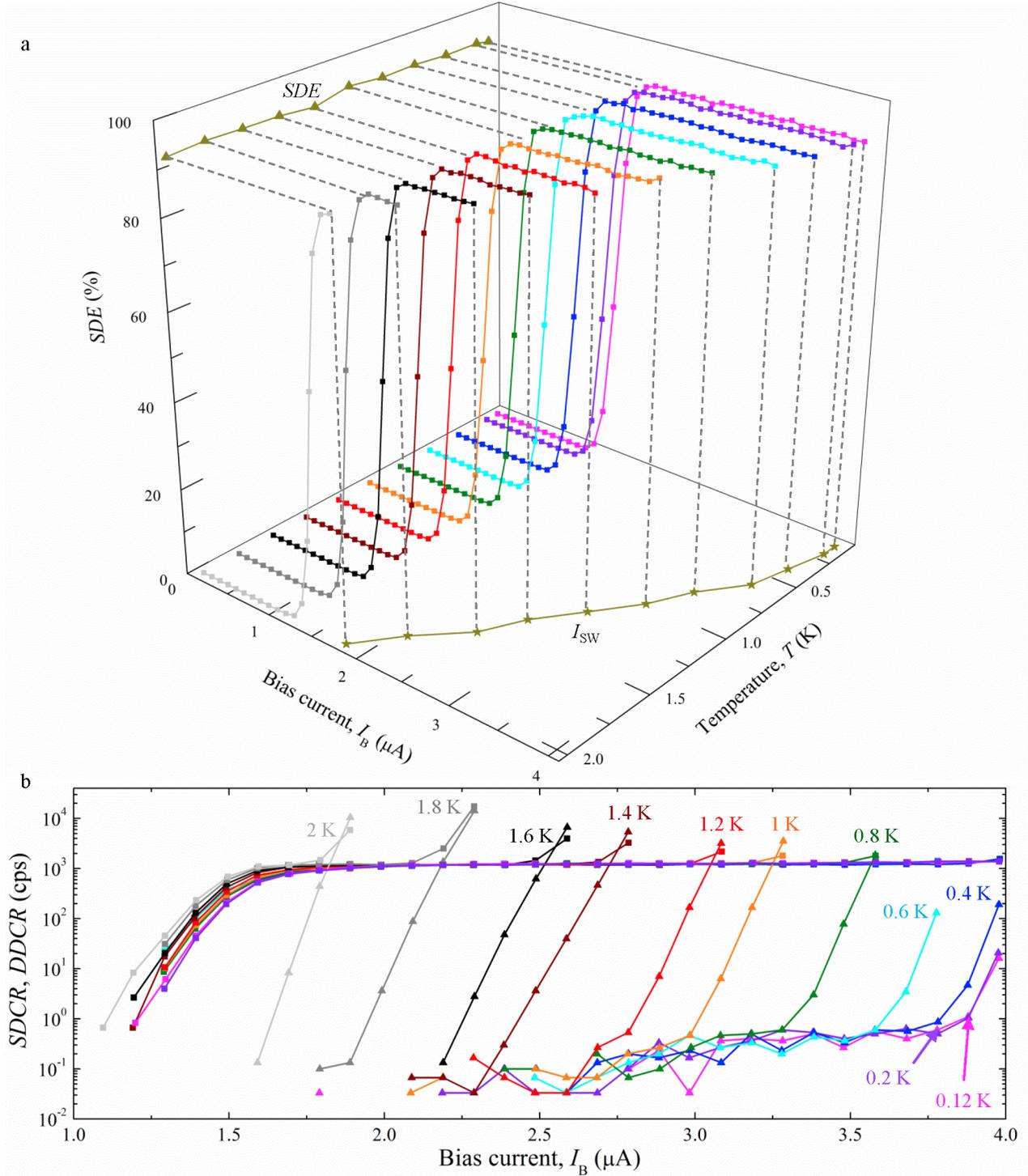

**Figure 3. Temperature dependence of the *SDE*, of the *SDCR*, and of the *DDCR*. a** *SDE* vs bias current ($I_B$) and temperature (*T*). The dark yellow curves on the $I_B$ - *T* and *SDE* - *T* planes represent the temperature dependence of $I_{SW}$ (stars) and of the *SDE* at $I_B = I_{SW}$ (triangles). **b.** *SDCR* (squares) and *DDCR* (triangles) vs bias current ($I_B$) in the temperature range *T* = 0.12 − 2 K. The *SDCR* and *DDCR* curves were obtained by averaging 3 subsequent acquisitions of the curves.

We characterized the timing performance of the detector system by measuring the histogram of the interarrival time of the response pulses (the time between two consecutive response pulses when the SNSPD was illuminated with a CW laser) and the timing jitter of the SNSPD. Although in conventional NbN SNSPDs the decay time of the response pulse has been traditionally used as an estimate of the reset time of the detector[25], in our detector the reset time was significantly shorter than the decay time. As shown in Figure 4a, the decay time of the response pulse of the SNSPD ($\tau$, which we defined as the time required for the pulse to decay from 90% to 10% of the maximum of the pulse) was $\tau \approx 120$ ns. However, as Figure 4b shows, the reset time of the detector ($t_R$, which we defined as the time at which the histogram of the interarrival time reached 90% of its peak value[21,26]) was as low as $t_R = 40$ ns. The fact that $t_R$ was a factor of $\approx 3$ lower than $\tau$ was due to the low $I_{co}$ of the detector ($I_{co} \approx 0.4 I_{SW}$, see Figure 1a). Indeed, when the SNSPD switched back to the superconducting state after a hot spot nucleation event, it was sufficient that the current in the nanowire increased above $\approx 0.4 I_{SW}$ for the *SDE* to recover fully. Figure 4c shows the instrument response function (IRF) of the detector system illuminated with a femtosecond-pulse laser for two different bias currents. The IRF became broader with decreasing $I_B$. Figure 4d shows the current dependence of the jitter of the detector system, which we defined as the FWHM of the IRF. The system jitter decreased from 250 ps at $I_B = 0.67 I_{SW}$ to 150 ps at $I_B = 0.97 I_{SW}$. The jitter of our detector system was higher than the values of 30 - 50 ps typically reported for conventional NbN SNSPDs[27]. We attributed the higher system jitter to the electrical noise of the read out circuit, rather than to the possibility that WSi SNSPDs had larger intrinsic jitter than NbN SNSPDs (see Supplementary Information).

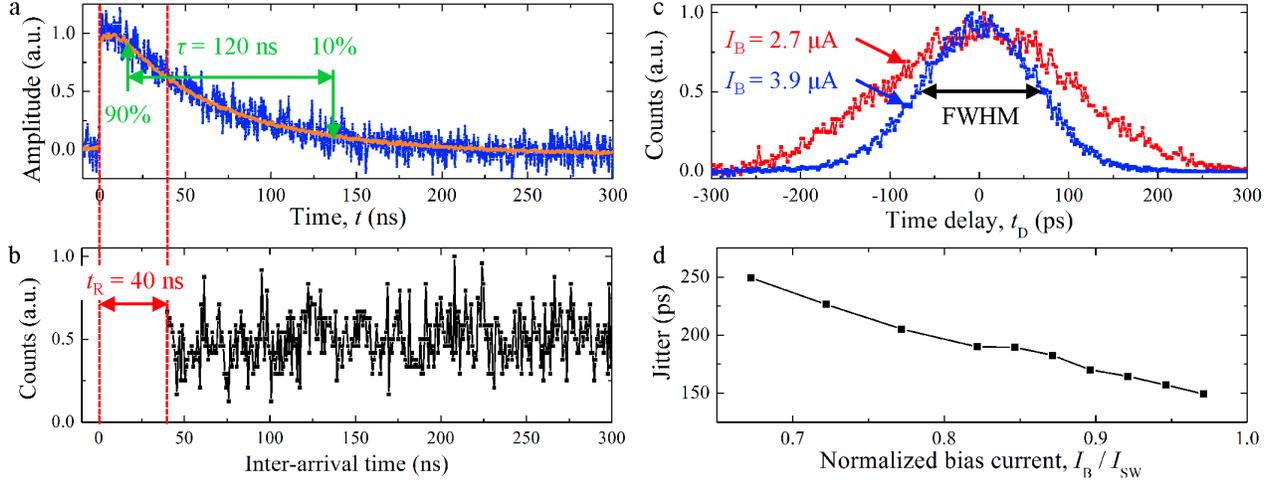

**Figure 4. Reset time and jitter. a.** Single-shot (blue curve) and averaged (orange curve) oscilloscope traces of the response pulse of the SNSPD biased at $I_B = 3.8$ μA ($I_B = 0.95 \, I_{SW}$). The time ($t$) at which the curves reached 50% of the maximum of the average trace (158 mV) with positive slope was set to $t = 0$ s. The curves were normalized by the maximum of the average trace. The vertical green arrows indicate the time at which the decaying part of the pulse reached 90% (at $t = 18$ ns) and 10% (at $t = 138$ ns) of the maximum. **b.** Interarrival time histogram of the SNSPD biased at $I_B = 3.8$ μA ($I_B = 0.95 \, I_{SW}$). The histogram was normalized by its maximum value. **c.** Instrument response function (IRF) of the SNSPD biased at $I_B = 3.9$ (blue curve) and $I_B = 2.7$ μA (red curve). The IRF at a particular $I_B$ was obtained by calculating the histogram of the time delay ($t_D$) between the rising edge of the synchronization pulse of the laser and the rising edge of the response pulse of the SNSPD. Each IRF was normalized by its maximum value. The black arrow indicate the FWHM of the IRF acquired at $I_B = 3.9$ μA. **d.** Current dependence of the jitter of the detector system. The data shown in Figure a to d were acquired while the detector was at a temperature of $T = 120$ mK, and the switching current of the SNSPD was $I_{SW} = 4$ μA.

In conclusion, our single-photon detector system based on WSi SNSPDs showed system detection efficiency $SDE > 90\%$ around $\lambda = 1550$ nm and device dark count rate $DDCR < 10$ cps up to a temperature of $T = 2$ K. We expect our detector system to achieve a system dark count rate limited by the device intrinsic dark count rate ($SDCR \approx DDCR < 1$ cps) by filtering the blackbody photons with a cold filter. In the future, by adopting a parallel architecture (superconducting nanowire avalanche photodetector, SNAP[21,28,29]), we expect to reduce the reset time of our SNSPDs below 10 ns and to increase the signal-to-noise ratio[21], which would allow the jitter of the detector system to be reduced.

Finally, because of the relatively large bias range with saturated detection efficiency at $\lambda = 1550$ nm, WSi SNSPDs have the potential for high fabrication yield across a silicon wafer and broad wavelength sensitivity[17,30]. These two features will enable two major advancements in the near future: (1) high system detection efficiency in the mid-infrared wavelength range; and (2) large SNSPD arrays with near-unity efficiency from the visible to the mid-infrared spectral regions.

**Acknowledgements**

We thank Dr. R. M Briggs, Dr. S. D. Dyer, Dr. W. H. Farr, Dr. J. Gao, Dr. M. Green, Dr. E. Grossman, Dr. P. D. Hale, R. W. Leonhardt, Dr. I. Levin, and R. E. Muller, for technical support, and Dr. S. Bradley, B. Calkins, Dr. A. Migdall, and Dr. M. Stevens, for scientific discussions.

**Methods**

**Detector system and measurement set up**

The experimental set up used for the optical characterization of our detector system is shown in Supplementary Information. For the system detection efficiency and the interarrival time measurements, we illuminated the detector with a fiber-coupled continuous-wave tunable laser with tuning range $\lambda = 1510 - 1630$ nm. For the jitter measurements, we used a mode-locked fiber laser with emission around 1560 nm, <100 fs pulse width, and $\approx 35$ MHz repetition rate. We controlled the polarization of the light from the lasers with a polarization controller. The light was then coupled to three variable optical attenuators (with nominal attenuation $A_1$, $A_2$, and $A_3$) and to a MEMS (micro-electro-mechanical system) optical switch. The optical switch diverted the light at its input to the detector system (we call this output *detector port*) or to a calibrated (see Supplementary Information) optical InGaAs power meter (we call this output *control port*).

After fabrication, a device could be removed from the wafer[19] and mounted inside a zirconia sleeve that held an optical fiber. Holding both the detector chip and the optical fiber, the zirconia sleeve realized the optical alignment with a typical accuracy of $\pm 3$ µm[19]. All of the optical fibers used were silica C-band single-mode fibers. The optical fiber coupled to the detector inside the cryostat was coated with a multi-dielectric-layer anti-reflection coating (ARC), which reduced the reflectivity ($\rho$) at the interface between silica and air (or vacuum) below 0.3% in the wavelength range of interest. The fiber coupled to the detector was then spliced to a fiber inside the cryostat. That cryostat fiber was fed out of the cryostat through a vacuum feed-through and then spliced to a fiber coupled to the detector port of the optical switch.

The detectors were wire bonded to launching pads connected by SMP connectors to brass coaxial cables (2 GHz electrical bandwidth at 300 K). The devices were current-biased with a low-noise voltage source in series with a 10 kΩ resistor through the dc port of a room-temperature bias-tee (40 dB isolation, 100 kHz – 4.2 GHz bandwidth on the RF port). The read-out circuit consisted of a chain of two low-noise, room-temperature amplifiers (100 kHz - 500 MHz bandwidth, 24 dB gain, 2.9 dB noise figure) connected to the RF port of the bias-tee. The amplified signal was connected to a 225 MHz-bandwidth counter (for detection efficiency measurements), or to a 8 GHz-bandwidth, 20 Gsample/s oscilloscope (for jitter and interarrival time measurements).

**Estimation of the system detection efficiency**

The *SDE* was measured as the ratio of the photoresponse count rate (*PCR*) and the number of photons in the SNSPD fiber ($N_{ph}$): $SDE = PCR / N_{ph}$. *PCR* was estimated as the difference between the response-pulse count rate (*CR*), measured with the laser beam attenuated $\approx 80$ dB ($A_2 = A_3 = 40$ dB) and coupled to the detector, and the system dark count rate (*SDCR*). We defined the *SDCR* as the response pulse count rate measured with the laser beam blocked by the shutters of the variable optical attenuators. $N_{ph}$ at a particular wavelength ($\lambda$) was calculated by using an estimate of the optical power in the SNSPD fiber ($P_{SNSPD}$) and the energy of a single photon at that wavelength.

We measured the *SDE* at a particular wavelength using the following procedure: (1) we measured the splitting ratio of the optical switch ($R_{SW}$), which we defined as the ratio between the power at the detector and control ports of the switch; (2) we measured the real attenuation of attenuator 2, 3 ($a_{2,3}$) when the nominal attenuation of attenuator 2, 3 was set to 40 dB ($A_1 = A_{3,2} = 0$ dB and $A_{2,3} = 40$ dB); (3) with the attenuation of attenuator 2 and 3 set to zero ($A_2 = A_3 = 0$ dB), we varied the attenuation of attenuator 1 ($A_1$) to obtain the desired input optical power in the control port ($P_C$); (4) we closed the shutters of the three attenuators and measure the *SDCR* vs $I_B$ curve; (5) we opened the shutters of the three attenuators, set the attenuation of attenuator 2 and 3 to 40 dB ($A_2 = A_3 = 40$ dB) to reduce the optical power to the single-photon level ($\approx 50 \cdot 10^3$ photons per second), and measured the *CR* vs $I_B$ curve. We calculated the optical power in the SNSPD fiber as $P_{\text{SNSPD}} = P_{\text{C}} \cdot \alpha_2 \cdot \alpha_3 \cdot R_{\text{SW}} / (1 - \rho)$. Further details are reported in Supplementary Information.